# Orientation-related twinning and dislocation glide in a Cantor High Entropy Alloy at room and cryogenic temperature studied by *in situ* TEM straining

Daniela OLIVEROS[1,2], Anna FRACZKIEWICZ[3], Antonin DLOUHY[4], Chen ZHANG[5], Hengxu SONG[5], Stefan SANDFELD[5], Marc LEGROS[1]

1 - CEMES-CNRS, 29 rue J. Marvig, 31055 Toulouse France
2 - Université de Toulouse, 118 route de Narbonne, 31400 Toulouse France
3 - MINES Saint-Etienne, Université de Lyon, CNRS, UMR 5307 LGF, Centre SMS, 42023, Saint-Etienne, France
4 - Institute of Physics of Materials and CEITEC IPM, Czech Academy of Sciences, 616 62 Brno, Czech Republic
5 - Forschungszentrum Jülich GmbH, Institute for Advanced Simulations: Materials Data Science and Informatics (IAS-9), 52425 Jülich, Germany

**Abstract**: *In situ* straining experiments were performed in a TEM on an equimolar CoCrFeMnNi (Cantor) high entropy alloy at room and cryogenic temperature. Perfect and partial dislocation activity were recorded in both cases. Twinning directly follows the development of partial dislocation shearing that has various origins (perfect dislocation splitting, anchoring). It is shown that, although twinning is more frequently observed at liquid nitrogen temperature, its prevalence depends mainly on crystal orientation. As a result, twinning and perfect dislocation plasticity are likely to occur jointly in random oriented polycrystals, even at early stages of deformation.

**Introduction**
High Entropy Alloys (HEA) were born from the idea that regular alloys constituted of one or two main elements represented a very small fraction of the combinatorial possibilities offered by the many metallic elements of the periodic table [1,2].
Their name was derived from the early belief that the configurational entropy of mixing many different elements would overcome the enthalpy gain of stabilizing intermetallic phases.
This concept has been challenged since then [3,4], and although other names, may be more relevant, have been suggested (MPEAs- for Multiple Principal Element Alloys), many authors still rely on the term HEA to call this new class of metallic alloys.
Since the early 2000s, a considerable body of literature has emerged to explore their processing and properties. In most recent review articles, e.g. [5], many poorly understood properties of these alloys are exposed and pave the way for future studies. As for other HEAs or MEAs (Medium Entropy Alloys containing 3 or 4 principal elements) of fcc structure, the mechanical properties of the CoCrFeMnNi (Cantor) alloy and its family are remarkable [6]. In particular, their toughness remains very high at cryogenic temperatures [7]. The principal hypothesis suggested for this intriguing property in literature [7-11] is the propensity of fcc-structured HEAs to twin at low temperatures and up to 500°C [8]. However, the few in-depth studies on the subject give very partial answers as only a few testing conditions were reported, even if a potentially large number of influential parameters (temperature, orientation, strain, stacking fault energy...) exist [9–13].

In fact, the mechanical properties of fcc metals rely heavily on the behavior of the main dislocation slip systems that, in a vast majority of deformation conditions, consist of a/2<110> type dislocations gliding on (111)-type planes. These dislocations are asymmetrically split into two a/6<112> dislocations separated by a stacking fault ribbon, the width of which is inversely proportional to the stacking fault energy (SFE) [14]. In high-SFE metals such as Al, the separation of both partial is very small, favoring the cross-slip of perfect dislocations and preventing twinning in most of the cases [15]. On the contrary, in low-SFE metals such as Cu, Ag [16] the large splitting of perfect dislocations into partials favors planar glide, extended stacking fault development and twinning. This can be beneficial as it offers both additional deformation modes and hardening possibilities, but may also lead to structural instability via martensitic transformation as in low-SFE steels [17].

In HEAs, the situation is more complex. The fact that the position of the various chemical species is random (including in the slip plane) precludes any easy prediction of the SFE. Recent Transmission Electron Microscopy (TEM) observations of dissociated dislocations lead to SFE estimations in the range of 18-27mJ.m$^{-2}$ [18], but larger fluctuations have been observed and may extend this range [19]. These values come from static observations. Overall, the proportion of twinning versus perfect dislocation glide remains a question as both systems have been observed across a wide temperature range, and the mechanisms by which these two kinds of plasticity unfold in fcc HEAs are still questioned. The purpose of this study is an in-depth characterization of the mechanisms of plastic deformation, on the basis of an extensive *in situ* TEM study, with an analysis of more that 60 different grain orientations. The study was performed on an equimolar CoCrFeMnNi fcc HEA (Cantor alloy) both at room (RT) and liquid nitrogen (LN$_2$) temperatures. We show that low temperature favors twinning but that grain orientation is critical to promote this type of plasticity.

**Material and Methods**

Ingots of CoCrFeMnNi equimolar alloy (Cantor alloy) were prepared from pure metals pellets or powders (purity always higher than 99,9% purity) following two methods. After cold crucible melting and homogenization (MINES Saint-Etienne), the ingots were hot forged and annealed for 2 hours at 1000°C under vacuum [20]. The second batch of samples were manufactured by arc melting (8.4x10$^{-4}$ Pa Ar atmosphere) and drop casting at ORNL and provided by the Institute of Physics of Materials in Brno. The Mn weight loss during arc melting is compensated by adding 1g (for a 475 g ingot) and a Zr pellet was melted before to clean the chamber from residual oxygen. Arc-melted buttons were flipped and re-melted 5 time for homogenization, drop cast in rectangular Cu molds, and solution-annealed 48h at 1200°C. After being cold-rolled from 12.7 mm down to 4 mm thick slabs, they were annealed 1H at 900°C. For processing details see [21].

Both processes resulted in a homogeneous single phase microstructure, with large grains (≈50 μm). The lowest dislocation density was observed in the fully recrystallized alloy (IPM, Brno), although no systematic measurements were performed due to grain-to-grain variations.

Rectangular (3 x 1.5 x 0.5 mm) samples were cut from these ingots and slabs using electro-discharge machining. They were mechanically thinned down to about 30 μm with SiC papers.

Final electropolishing was made in a Struers Tenupol twin-jet polishing unit using a 10%Perchloric acid-90% Ethanol electrolyte to create thin regions around a central hole (Fig. 1a). X-ray Energy Dispersive Spectroscopy (EDS) was performed on the thicker part of the polished zone, and within the accuracy of the technique (1-2% without the use of standards) the equimolar composition of samples was verified.

These rectangular samples are then glued with cyanoacrylate onto Cu grids specially designed for Gatan straining holders (Fig. 1c). In this study, a cryo-straining holder served both for liquid nitrogen (LN$_2$) and room temperature (RT) *in situ* experiments. The so-called "LN$_2$ experiments" were actually performed at 100-105K, as the sample is cooled down through a cold finger running through the TEM holder and connected to its LN$_2$ reservoir. 100K is close to the minimum reachable with this holder, with some fluctuations depending on the outside temperature and the level of vacuum reached in the reservoir shell. We used it without temperature regulation, at the minimum attainable temperature for a given experiment. More details about sample preparation and *in situ* straining experiments can be found in videos posted online [22,23].

Finite Element Method (FEM) calculations made with CATIA® illustrate the fact that the thin regions along the straining axis are those where the stress reaches a maximum (Fig. 1b). The FEM calculation is using a linear elastic material model. Therefore, all forces and stresses are linearly related and can therefore be easily scaled to larger elastic stresses.

Moreover, the resolved shear stress (RSS) τ acting on dislocations moving during in situ tests was also estimated as the line tension stress:

τ=μb/R, where b is the Burgers vector and R the radius of curvature of the dislocation. The shear modulus μ of the Cantor alloy at RT and LN2 temperature was taken from the work by Haglund *et al.* [24].

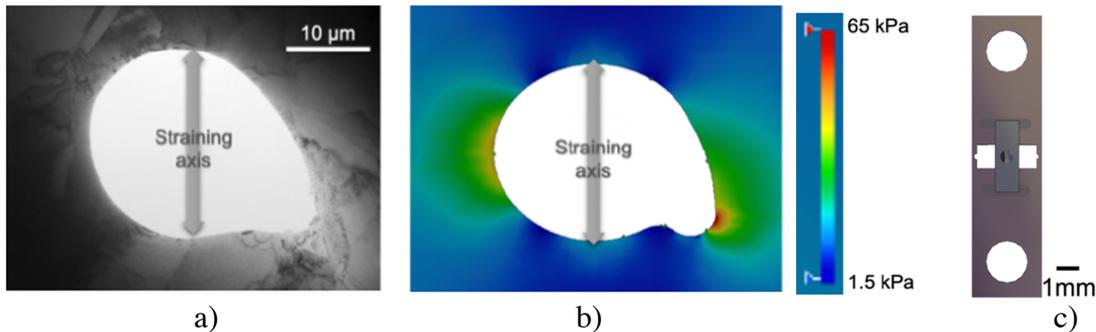

*Figure 1. Typical shape and stress concentration around in the electron-transparent zone in an electropolished sample for in situ TEM straining. a) Low magnification TEM image, b) FEM stress concentration in the thin regions around the hole (FEM calculation made with CATIA V5.6, applied force of 90N, Min (blue)/Max (red) Von Mises Stress = 1.5 / 65 kPa), c) sample glued on a straining Cu grid.*

Under *in situ* deformation, orientation of the active grain was determined by capturing 3 to 4 two-beam diffraction patterns (DPs) using the single tilt function of the TEM (a JEOL 2010 LaB$_6$ operating at 200kV). DPs, micrographs and videos were all recorded using a Megaview III camera from Soft Imaging System (now EMSIS) and stored on hard drives. Unless stated otherwise, the straining axis is vertical in all the micrographs and videos (Figure 2a), corresponding to the direction going from pole to pole on the stereographic projection (Figure 2b) obtained from the captured diffraction patterns (insert in Fig. 2a). This combination allows us to directly index the activated slip planes (Fig. 2a, c) and, subsequently, the local thickness of the TEM foil (Fig. 2d). PycoTEM, an open source software developed at CEMES, was also used to construct these stereographic projections [25]. During in situ experiments, short strain pulses (on the order of $10^{-3}$ s$^{-1}$) are applied, separated by longer periods where the dislocation movements are observed. In average, the strain rate of the in situ experiments are in the order of $10^{-4}$ s$^{-1}$ to $10^{-5}$ s$^{-1}$.

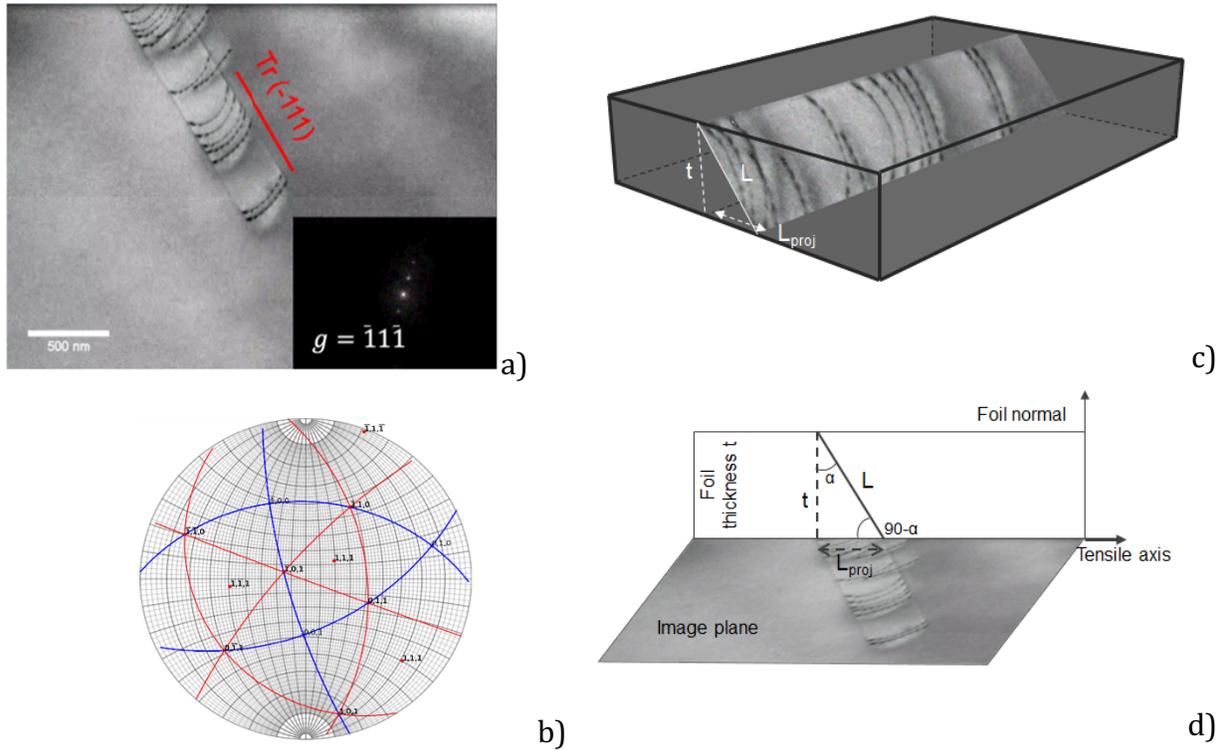

Figure 2. Orientation and foil thickness measurement during an in situ TEM straining test. a) Dislocation pile-up in a grain and corresponding Diffraction Pattern (DP) on the lower right. b) Corresponding stereographic projection (111)-type planes are in red), the activated plane is (-111) in this case. Straining axis is vertical in a) and goes through both north and south poles of the stereographic projection of b). c) Sketch of the dislocation pile-up in its (-111) glide plane. d) projected image and measured film thickness: $t = L_{proj}.\tan(90-\alpha)$. In this relation L_proj is the same as in Fig. 2c only for small sample tilt angles which is true for our setup. In this example, $L_{proj}=430$ nm, and=25°, so t = 200 nm.

**Results**

About 25 *in situ* TEM straining experiments were carried out at room and LN$_2$ temperatures, representing more than 60 different grain orientations. Strain increments are imposed on the sample through the motor of the straining holder and observations are made during subsequent relaxation time. Dislocation activity initiates in stressed regions of the foil (Fig. 1c) and from all the usual stress-concentration sites: grain boundaries (GBs), entanglements of sessile dislocations, thin foil edges. This generates localized plasticity that subsequently unravels in the sample.

***Low temperature plasticity: LN2 temperature experiments***

Figure 3 illustrates the beginning of an *in situ* TEM straining experiment at 100K. First, the applied stress mobilizes dislocations that are left over from a denser microstructure in the annealed alloy (top of Fig. 3b, c). This leads to an intra-granular, distributed plasticity. When dislocations can easily cross a GB, plastic deformation may remain spread, but this happens rarely. In any opposite case, activation of dislocation sources (in thin foils, only single spiral sources are observed) generally lead to long and dense pile-ups of dislocations (as visible in Fig 3 and Fig 6) that concentrate the deformation along a restricted set of (111)-type planes. It can be seen that the pile-up in figure 3 densifies after an increase in strain, and that the dislocations' radius of curvature diminishes (e.g. stress increases), even for those dislocations that are situated at the end of the pile-up. Curvature measurements made on the last moving dislocation in this pile-ups leads to a Resolved Shear Stress (RSS) of 53 MPa, corresponding to an applied stress of 112 MPa. Such local stress measurement underestimates the overall applied stress as it also incorporates the back stress of the pile-up and a possible friction stress. In figure 3b, a dislocation followed by a stacking fault (SF) is emerging from the back of the pile-up. This partial dislocation is followed by others coming from the thinner part of the foil (right side in Fig. 3), with the same Burgers vector. In figure 3c, the characteristic grey contrast (with fringes) in regions 1, 2, 3 attests that the initial perfect dislocation pile-up is being replaced by a micro-twin (which implies that partial dislocations lie on adjacent planes rather than on the exact same one). The way this transformation occurred is not yet identified, but the reaction seen in Figure 3c, indicated by "R", where perfect dislocations (coming from the left) dissociate at the interaction with a sessile dislocation that serves as obstacle, and form micro twins (right) is commonly observed.

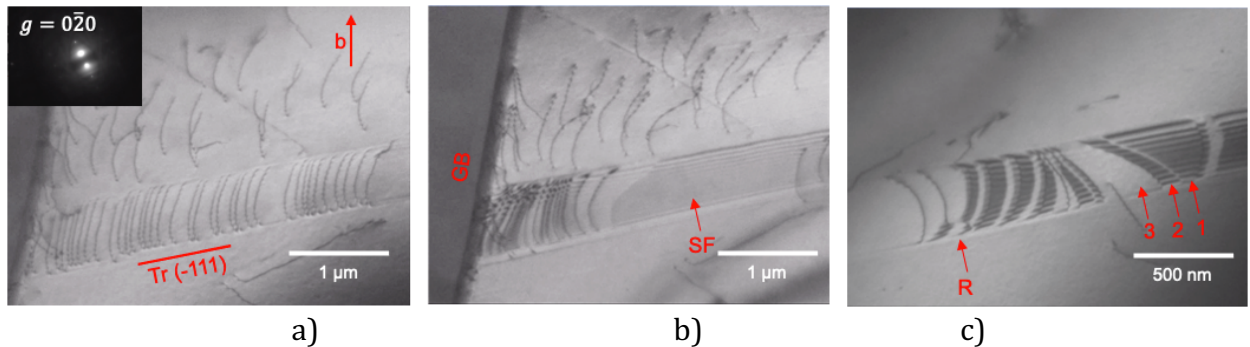

a) b) c)

*Figure 3. Development of plastic deformation during an in situ TEM experiment performed at 100K. a) Perfect dislocation with b=a/2[0-11] and gliding on the (-111) plane blocked by a GB. b) After a strain/stress increment, dislocations move towards the left where they are blocked by a grain boundary (GB). A partial dislocation, leaving a stacking fault (SF) is incorporated in the pile-up. c) Upon unloading, and in a thinner part of the sample, the back of the pile-up is constituted by partial dislocations (b= a/6[-1-12]) forming a micro-twin. Each image is taken ≈30 minutes after the preceding one.*

The development of plasticity at $LN_2$ temperature often involves a mix of perfect dislocations and twinning.
Figure 4 shows more snapshots extracted from an *in situ* TEM straining experiment carried out at 104K where two pile-ups of perfect dislocations located on (-1-11) and (-111) planes are visible. Perfect dislocations in these pile-ups are a/2[011] (Schmid Factor= 0.02) and a/2[-1-10] (Schmid Factor = 0.46) respectively, which means that the later will be much more favored than the former when applying an external stress (always along the vertical axis on the micrographs). One widely split dislocation separated by a stacking fault is visible in the (-

111) plane in figure 4a. After applying a 2μm displacement increment on the holder, plastic deformation intensifies on (-111) plane mainly by the propagation of partial dislocations that leave longer stacking faults (Fig. 4). The pile-up on (-1-11), mainly formed by perfect dislocations, does not move because its Schmid factor is too low. On the contrary, perfect dislocations moving on (-111) are more active (large Schmid factor), as for partial dislocations (b=a/6[211] Schmid Factor = 0.47) that pull stacking faults under stress. As seen in figure 4b, the inactive perfect dislocations on (-1-11) serve as obstacles to the slip of a/6[211] partial dislocations on (-111). The darker contrast visible in the middle of the stacking fault is due to a partial dislocation gliding towards the (-1-11) pile-up close to a previously drawn stacking fault. Once a third partial follows the same path, a micro-twin is created, provided the partial dislocations glide on adjacent planes and have the same Burgers vector.

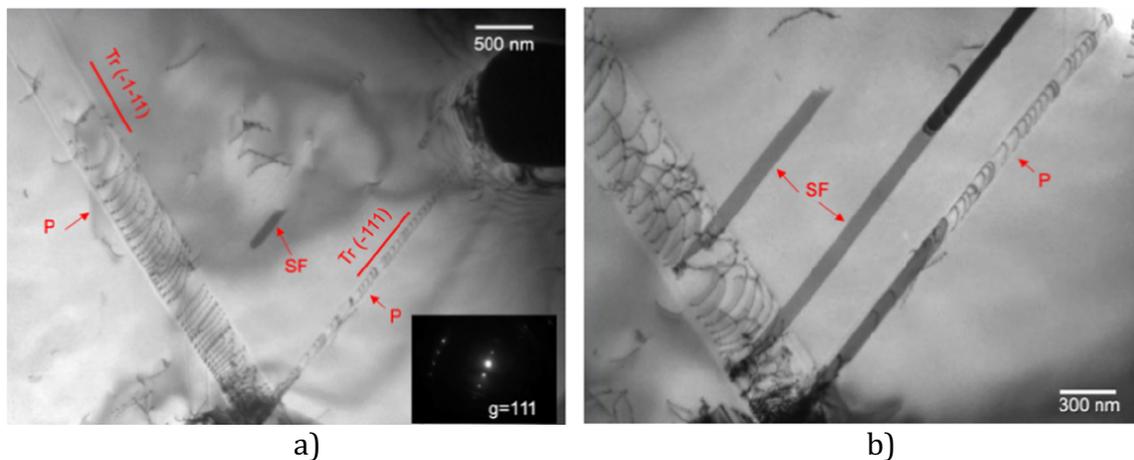

*Figure 4. Mix of perfect and partial dislocations in the development of plastic deformation during an in situ TEM experiment performed at 104K in a part of the foil where the thickness is 165nm. a) Low magnification of a zone deforming plastically., containing perfect dislocations pile-ups (P) and one widely split dislocation separated by a stacking fault (SF). b) Upon straining, glide of partial dislocations on (-111) leaves longer stacking faults. Note that dislocation lines are in general more cusped than at room temperature.*

### *Room temperature plasticity*

At room temperature, incipient plasticity can also develop through partial dislocation activity and twinning. Figure 5 shows several snapshots extracted from an *in situ* TEM straining experiment carried out at 300K.

Each "dark triangle" in figure 5a is a dissociated a/2 [011] dislocation gliding on the (-1-11) plane. As the applied strain (and thus stress) increases, the dissociation expands too (Fig. 5b, c). The process appears reversible as a decrease of the applied stress causes the dissociation to shrink back almost to the initial configuration (Fig. 5d). Continuing the straining experiment spurs the extension of the stacking faults separating both partials to the point where a recombination becomes impossible. Leading and trailing partial dislocations interact with other defects (dislocations on other glide planes, GBs) and their separation distance increases until their elastic interaction is negligible. The network formed by these stacking faults on two different sets of glide planes (Fig. 5e, f) constitutes a large array of potential obstacles for other dislocations, perfect or partial.

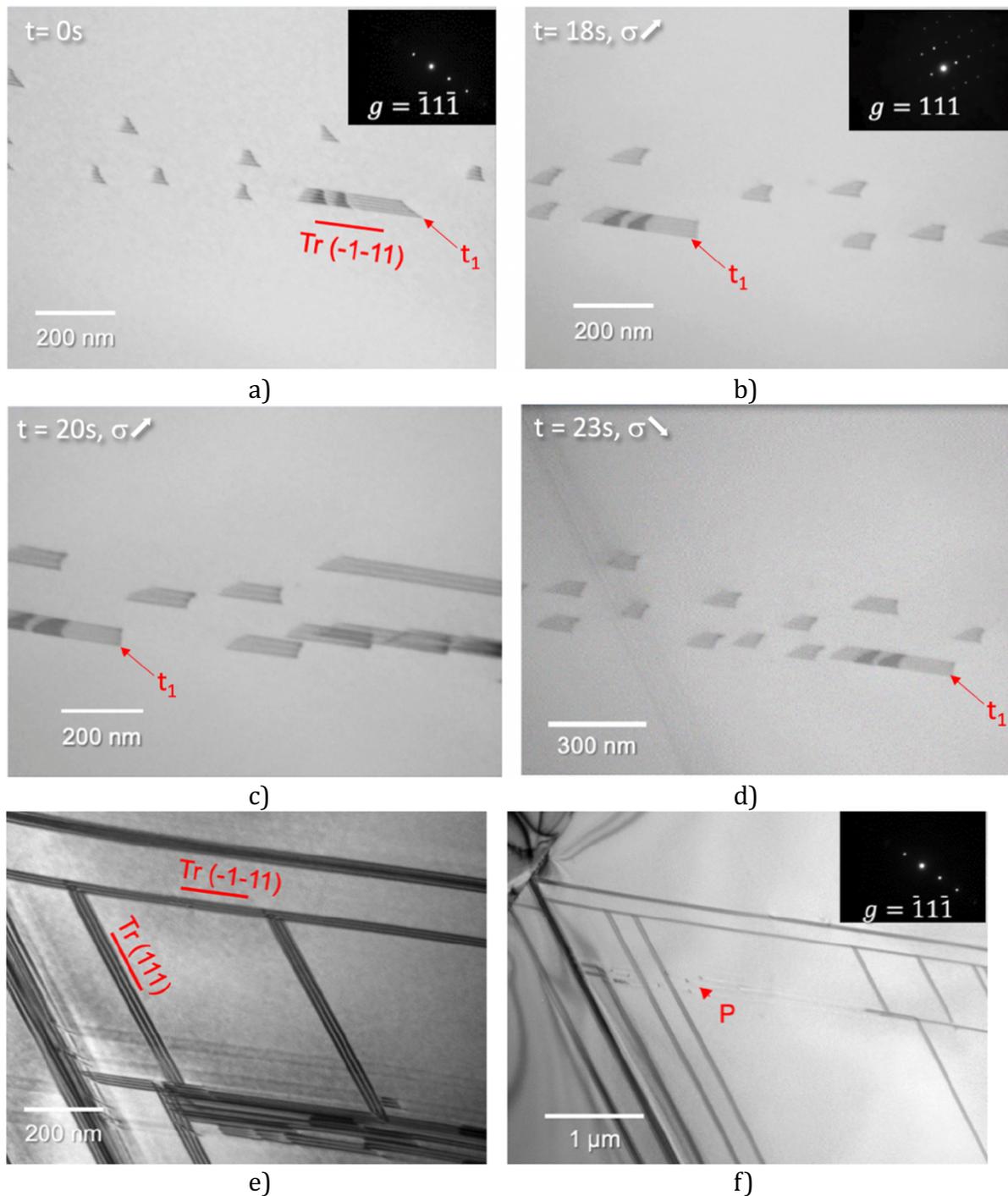

*Figure 5. Splitting of perfect dislocation leading to extended stacking faults during a 300K in situ TEM experiment. a)-c) Swelling dissociation under increasing strain (taken at t=0, 18' and 20' respectively). d) Shrinking dissociation under decreasing strain (t=23'). $t_1$ shows the same trailing partial dislocation. e) Development of extended stacking faults, easily recognizable by their dark grey fringe contrast, on (-1-11) and (111) planes during the experiment. f) After the experiment and the release of the strain, most of the extended stacking faults shrunk and disappeared. A few perfect dislocations are also visible (P).*

These long stacking faults ribbons and individual partials lead to micro-twinning upon further straining. The transformation of perfect dislocation into micro-twins and inversely has also been observed after reactions on obstacles (R in Fig. 3), usually forest dislocations. Again, we did not observe a clear transition from one deformation mode to the other related to the amount of strain (density of dislocation increasing nor stress concentration due to strain hardening). Twinning can occur right at the onset of plasticity or after a few percent of deformation (typically, *in situ* straining tests do not extend beyond 10% strain).

At room temperature, perfect dislocations are however more often observed. Figure 6 shows a typical set of perfect a/2 <110> dislocations gliding on (1-11) planes. Dense pile-ups develop against strong obstacles such as GBs. As for low temperature experiments, we also measured dislocation curvatures on the last dislocation of the pile-up and this returned a much lower value of the RSS: 22 MPa in this case, corresponding to 55MPa of applied stress. Again, this value is underestimated as it does not take into account the back-stress from the first dislocations of the pile-up nor possible friction stress. Nonetheless, it is worth noting that this stress value, measured in similar conditions at low and room temperature, reproduces well the macroscopic observations: in Cantor's alloy, the stress needed to deform the alloy increases when the temperature decreases into the cryogenic area.

It can also be seen in figure 6 that smaller obstacles, such as immobile dislocations are usually overcome by a few (less than ten) dislocations. Other "invisible" obstacles, such as impurities or short-range order chemical variations are also probably responsible for the cusps seen frequently on individual or collective dislocations (pile-ups and shear bands). These cusps impose much lower radii of curvature on dislocation lines (perfect and partial) and are observed much more frequently at low temperature than at room temperature. The strong curvature is an indication of the larger stress needed for the dislocations to overcome these obstacles. A quantitative and statistical study is currently carried out to assess the strength of the various obstacles present in this specific alloy.

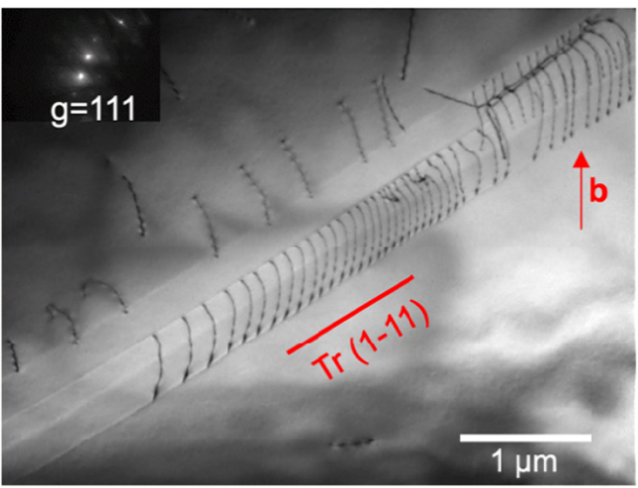

*Figure 6. Pile-ups of perfect dislocations during a 300K in situ TEM straining experiment. Both pile-ups are formed on two parallel (1-11) planes against a GB located further right and up in the sample (not visible in the picture) (b= a/2[011], (1-11) glide plane). Note that dislocation lines are more rounded (less pinning points) than at low temperature.*

As a result of the development of pile-ups, the deformation is concentrated in several shear bands where one or several (111) planes are heavily sheared. This concentration is partially lowered at GBs where some alternate glide planes are activated, unless a closely orientated slip system is available in the next grain.

We have summarized all these observations in figure 7. The presented graph identifies the crystallographic orientation of each tested grain in all the strained samples: each dot corresponds to the straining axis in a grain that has been deformed at room or $LN_2$ temperature. The pink (resp. green) areas regroup the grains in which only perfect dislocation glide (resp. only twinning) has been observed. The yellow area regroups the grains in which both modes of deformation co-existed during the straining experiment. It can be seen that both deformation modes are active at both temperatures, but that twinning is favored at low temperature. To be more precise, very few grains have demonstrated twinning-only deformation (green areas), but grains where both twinning and perfect dislocation activity co-exist (yellow area) represent a larger fraction of the orientations at low temperature (Fig. 7b). Inversely, at room temperature (Fig. 7a), a larger zone of the standard triangle is occupied by the pink cloud, corresponding to straining axis orientations leading to perfect dislocation deformation only. The same pink region is smaller at low temperature. In other words, for randomly orientated grains, observing perfect dislocations will be easier at room temperature.

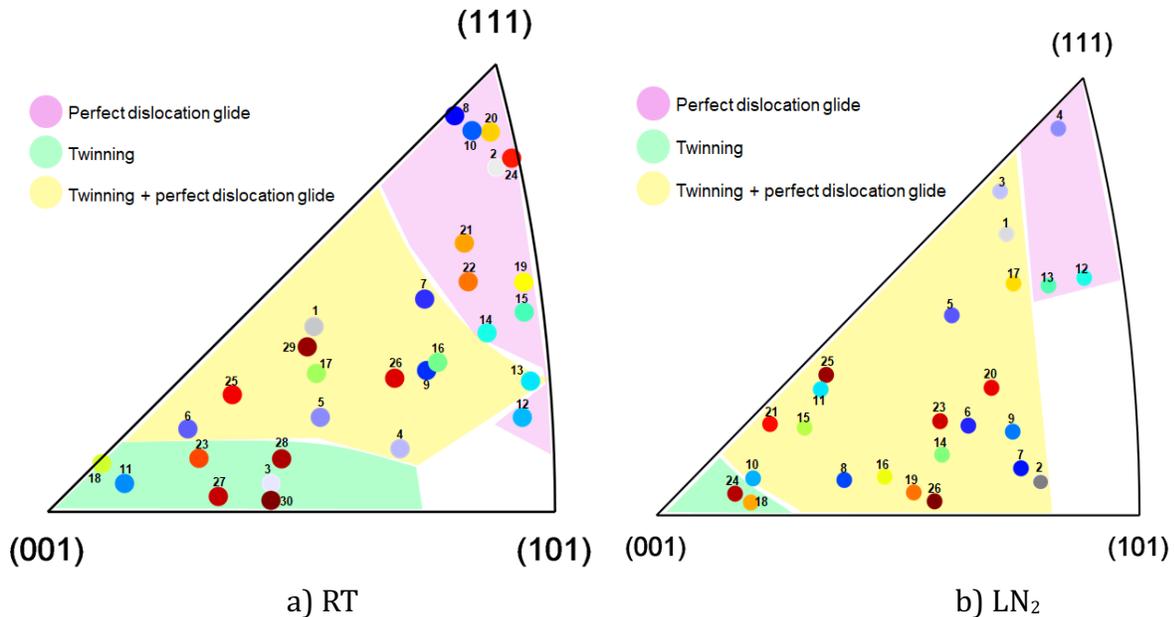

*Figure 7. Orientation dependence of micro-twinning vs perfect dislocation glide in standard triangles at room (a) and $LN_2$ (b) temperatures.*

**Discussion**

The present study underlines the fact that two modes of plasticity, mechanical twinning (including glide of isolated partial dislocations) and perfect dislocation glide, coexist in the equimolar CoCrFeMnNi alloy at both tested temperatures: around 100K ($LN_2$, liquid nitrogen) and 300K (RT, room temperature). The key parameter traditionally invoked to explain the capacity of a metal or an alloy to deform by twinning is the stacking fault energy (SFE). For instance, aluminum, characterized by a very high SFE (over 100mJ/m² [26–28]) does not twin unless very specific circumstances [15]. On the other hand, a low SFE favors twinning, such as observed in copper, brass,… [29, 30], but also in low SFE austenite in austenitic stainless steels from Fe-Cr-Ni system [31, 32].

Zaddach et al. [18], using ab-initio calculations, estimated the stacking fault energy for the equimolar CoCrFeMnNi alloy between 18 and 27 mJ.m$^{-2}$. These authors have also shown a strong variation of SFE upon deviation of chemical composition from the equimolar one [18]. Experimentally, average values were given around 30 mJ.m$^{-2}$ [33], but very large fluctuations of dissociated dislocations suggest that either this $\gamma_{sf}$ value varies along a given dislocation line or that a "local SFE " may have to be considered [19]. Fluctuations of SFE may arise from local variation of chemical composition that could itself be influenced by the presence of dislocations, although this has not been clearly evidenced yet, even from recent *in situ* straining experiments at room temperature [34].

From our experiments, it is clear that an applied shear stress is able to split the asymmetric dissociation of perfect dislocations (Fig. 8), even at or just before the onset of plastic deformation, as exemplified in figure 5.

This can be rationalized in terms of projected local stress state onto activated dislocations. Figure 8 reproduces the variation of the Peach-Kohler shear stress acting on both partial dislocations $b_{p1}$ and $b_{p2}$ that compose a perfect dislocation of Burgers vector **b** gliding on a (xy) plane. For a tensile test, the stress difference $\tau'_d$ between the shear acting on the edge component of each partial dislocations ($\sigma^1_{yz}$ and $\sigma^2_{yz}$.) will favor the constriction of dislocation b when positive or its dissociation when negative. This so-called Escaig stress is then plotted in a standard stereographic triangle for the most favorable slip system (highest Schmid factor). $\tau'_d$ reaches the strongest negative values (red zone) near the 001 orientation. In the blue region, $\tau'_d$ is positive and the applied stress tends to reduce the splitting [35].

Comparing this standard triangle to those of figure 7, it could be expected that stacking fault extension and then twinning would occur mainly in the red zone, near 001. This is roughly verified, with twinning only occurring near the 001 pole at both temperatures and perfect dislocation glide taking place mainly near the 111 axis. What is surprising here is the extent of mixed modes (twin + perfect dislocations) towards this 111 pole, especially at low temperatures.

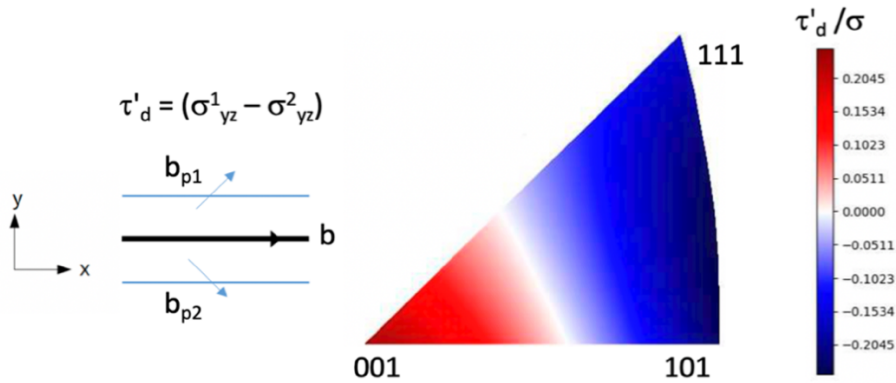

Figure 8. *Escaig split of two partial dislocations in function of the straining direction in tension. Left: $b_{p1}$ and $b_{p2}$ (a/6[112] type) compose a perfect dislocation b (a/2[110] type) split in the (xy) plane. $\tau'_d$ is the difference between the Peach Kohler stresses acting on the edge components of $b_{p1}$ and $b_{p2}$. Right: sign and amplitude of $\tau'_d$ in function of the direction of the applied stress in the standard stereographic triangle.*

A possible explanation for this extension of mixed modes lies in the local deviations to the macroscopic straining axis that occur in our thin foil configuration where a hole is punched in the center (Fig. 1). The straining axis has a tendency to align parallelly to the edges of the hole in the thinnest areas. Therefore, in a given experiment where several grains are stressed at once, we collect grain orientations where the actual tensile axis is within ±15° of the macroscopic one. As a consequence, reporting the macroscopic tensile axis in a given grain introduces an uncertainty regarding the actual Escaig stress in this grain where one mode should prevail over the other. This explains why the yellow zone in figure 7 (corresponding to an expanded region around the "neutral" white zone in figure 8) is much wider. As a consequence, the single mode zones (green and pink in Fig. 7) are reduced to areas very close to the 111 and 001 orientations.

The splitting of partial dislocations that takes place in figure 5 is emblematic of the mechanism predicted in figure 8. This grain however, labelled #16 in figure 7a, is not located in a splitting-favored zone of figure 8 (although close to the white separation line). Several reasons may explain this apparent discrepancy. As explained above, the straining axis may not be exactly vertical in this particular grain. The neighboring grains may also impose a deviation of the applied straining axis. The main reason is probably that the (-1-11) in which this dissociation occurs is not the one most favored by the Schmid factor. As Figure 8 only addresses the behavior of the preferred glide system (highest Schmid factor) we should re-draw the shear stress directions for this specific glide system, which is beyond the scope of this paper [35].

The fact that low Schmid factor slip systems are activated is not uncommon as the alloy, even in the annealed state, contains many remaining dislocations. These remnant dislocations can be activated even if their Schmid factor is not maximum, which seems to be the case here as their velocity under stress is much lower than in other grains.

Because of all these factors, concomitantly observing two modes of deformation (twinning and perfect dislocation glide) is not surprising. What our experiments show is that activation of a given system is largely influenced by the grain orientation with respect to the local straining axis, but that local factors such as the initial microstructure, may blur this picture and mix deformation modes in many cases. We also show that twinning is slightly favored at low temperature, but this is a second order effect.

Twinning is activated not only in zones where stacking faults were observed at the beginning of the deformation, but also after the reaction of perfect dislocations on forest-type obstacles (Fig. 3c). This observation reinforces the former hypothesis that twinning may develop once a sufficient density of obstacles exist (strain hardening, leading to a sufficient amount of stress [36]) or if the existing obstacles are oriented conveniently to create the right reaction with incoming dislocations (through cross-slip for instance [13]).

It can also be argued that working in a thin foil favors the splitting of dislocations as image forces act in opposite direction on both sides of the crystal [37]; thus, nucleation of partial dislocations can also be favored compared to perfect ones in thin foils [38]. Yet, both effects will influence the behavior of dislocations only in very thin foils (30-50 nm and below) which is far from our experimental conditions: most of our analyses were carried out in 100 to 300 nm thick regions. Also, as all experimental conditions were the same or very similar between all the samples that underwent *in situ* straining experiments (straining rate - typically in the range of $10^{-5}$ s$^{-1}$, mostly in relaxation mode, sample preparation…), the increased proportion of twins in samples tested at low temperature is indicative of a real trend.

A decrease of $\gamma_{sf}$ at low temperature would favor this effect, as anticipated by some authors from their ab initio calculations [39]. However, as seen before, not only simulations of this very complex class of alloys are to be validated by experiments, but those have revealed that variations of the $\gamma_{sf}$, because of local order or segregation effects can impact the dissociation by factors 2 or more.

In all the straining experiments, we observed that the pinning points arising from local chemical or order variations (different from forest/dislocation type) are stronger at low temperature. This feature is clearly evidenced by the fact that they impose steeper curvatures on mobile dislocations. As stated before, the motion of dislocation requires larger stresses at 100K (112 MPa vs 53 MPa, as measured on tail dislocations in moving pile-ups). Because these pinning points act similarly on partial or perfect dislocation, this action, combined with a low SFE will equally pin the leading and trailing partials of dissociated dislocations. We previously saw that a random orientation of the applied shear will favor the motion of the two partial dislocations of a dissociated perfect dislocation in the same direction (blue region in Fig. 8). This combined action of the stress on a pair of dislocations favors the overcoming of an obstacle. In the case where the stress acts to separate the dislocations, the stacking fault will pull the two partial dislocations together. In this context weak obstacles will not separate these partial dislocations. However, if obstacles become stronger, as it seems to be the case at low temperature, the probability that the trailing partial is retained by one of them increases. This will favor the dissociation and the development of long stacking faults (as in Fig. 4) and twins in the crystal as the deformation proceeds [6]. As demonstrated in our experiments, the presence of extended stacking faults favors the subsequent development of mechanical twinning. In contrast to the previous studies, we did not observe cases where twinning would be triggered by cross-slip [13], nor that a certain threshold of stress is required to activate this deformation mode [36]. In fact, the stress levels measured from the curvature radii of the mobile dislocations were very low, which is not surprising at the onset of plastic deformation. Even if these curvatures were measured in the tail of pile-ups where the back-stress of front dislocations reduces the local resolved shear stress, we are very far from the hundreds of MPa that are supposed to be needed to trigger twinning.

**Conclusions**

*In situ* TEM straining experiments were carried out at room and near liquid nitrogen temperature on a equimolar CoCrFeMnNi ("Cantor alloy") austenitic high entropy alloy. It was found that:
- Twinning and perfect dislocation glide occurred at both temperatures, and could be triggered at the onset of plasticity;
- The grain orientation with respect to the loading axis is the primary factor influencing the mode of deformation;
- In particular, the dissociation of perfect dislocation seems a common mechanism leading to the development of extended stacking faults. These extended stacking faults serve as seeds to the expansion of mechanical twins;
- The more frequent occurrence of twinning at low temperature could be the result of a lowering of the stacking fault energy and a more effective pinning of trailing partial dislocations.

This last point could be critical to explain the exceptional mechanical performances of such alloy at cryogenic temperatures. More experiments are underway to better characterize the type and strength of local pinning points in HEAs at low T.

**Acknowledgements**
This work was supported by the European Union's horizon 2020 research and innovation program under grant agreement n°823717 (ESTEEM3) and by the European Research Council through the ERC Grant Agreement No. 759419 MuDiLingo ("A Multiscale Dislocation Language for Data-Driven Materials Science").
José Daniel Orellana for CATIA FEM calculations.
Frédéric Mompiou for Escaig split stereographic calculations.
One batch of TEM samples was cut from CoCrFeMnNi slabs manufactured at Oak Ridge National Laboratory (Prof. E.P. George).